\documentclass[%
aip,jap,letterpaper,floatfix,
 preprintnumbers,
 reprint,
 amsmath,amssymb,
 showpacs,showkeys,
 superscriptaddress,
]{revtex4-1}
\usepackage{lineno,xcolor}
\usepackage{graphicx}
\usepackage{bm}
\usepackage[breaklinks=true]{hyperref}
\hypersetup{bookmarksnumbered, pdfpagemode=UseOutlines, 
pdfauthor={A.\ Aqeel}, 
pdftitle={Surface sensitivity of spin Seebeck effect}, pdfdisplaydoctitle, 
colorlinks=true, citecolor=blue, filecolor=blue, linkcolor=blue, urlcolor=blue}
\usepackage{enumitem}
\usepackage{epstopdf}
\usepackage{pdfpages}

\begin{document}

\renewcommand\linenumberfont{\normalfont\tiny\sffamily\color{gray}}

\title{Probing current-induced magnetic fields in Au$|$YIG heterostructure with low-energy muon spectroscopy}

\author{A.\ \surname{Aqeel}}
\affiliation{Zernike Institute for Advanced Materials, University of Groningen, Nijenborgh 4, 9747 AG Groningen, The Netherlands}
\author{I.\ J.\ \surname{Vera-Marun}}
\affiliation{Zernike Institute for Advanced Materials, University of Groningen, Nijenborgh 4, 9747 AG Groningen, The Netherlands}
\affiliation{School of Physics and Astronomy, The University of Manchester, Schuster Building-2.14, Manchester M13 9PL, UK}
\author{Z.\ \surname{Salman}}
\affiliation{Laboratory for Muon-Spin Spectroscopy, Paul Scherrer Institute, WLGA/U119, CH-5232, Villigen, Switzerland}
\author{T.\ \surname{Prokscha}}
\affiliation{Laboratory for Muon-Spin Spectroscopy, Paul Scherrer Institute, WLGA/U119, CH-5232, Villigen, Switzerland}
\author{A.\ \surname{Suter}}
\affiliation{Laboratory for Muon-Spin Spectroscopy, Paul Scherrer Institute, WLGA/U119, CH-5232, Villigen, Switzerland}
\author{B.\ J.\ \surname{van Wees}}
\affiliation{Zernike Institute for Advanced Materials, University of Groningen, Nijenborgh 4, 9747 AG Groningen, The Netherlands}
\author{T.\ T.\ M.\ \surname{Palstra}}
	\email[e-mail: ]{t.t.m.palstra@rug.nl}
\affiliation{Zernike Institute for Advanced Materials, University of Groningen, Nijenborgh 4, 9747 AG Groningen, The Netherlands}
\begin{abstract}
We investigated the depth dependence of current-induced magnetic fields in a bilayer of a normal metal (Au) and a ferrimagnetic insulator (Yttrium Iron Garnet - YIG) by using low energy muon spectroscopy (LE-$\mu$SR). This allows us to explore how these fields vary from the Au surface down to the buried Au$|$YIG interface, which is relevant to study physics like the spin-Hall effect. We observed a maximum shift of 0.4~G in the internal field of muons at the surface of Au film which is in close agreement to the value expected for Oersted fields. As muons are implanted closer to the Au$|$YIG interface the shift is strongly suppressed, which we attribute to the dipolar fields present at the Au$|$YIG interface. Combining our measurements with modelling, we show that dipolar fields caused by the finite roughness of the Au$|$YIG interface consistently explains our observations. Our results, therefore, gauge the limits on the spatial resolution and the sensitivity of LE-$\mu$SR to the roughness of the buried magnetic interfaces, a prerequisite for future studies addressing current induced fields caused by the spin-Hall effect.  
\end{abstract}
\maketitle
Recently the exciting field of spintronics has been transformed by new concepts to manipulate the spin transport taking place at the interfaces between magnetic and non-magnetic materials~\cite{vlietstra_simultaneous_2014,bauer_2012_nmat}. Therefore, it is important to understand the spatial distribution of spin accumulation in different devices.  The spin accumulations at these interfaces have been mostly created electrically by the spin-Hall effect (SHE) by sending a charge current through normal metal (NM) with strong spin-orbit coupling~\cite{Kato2004,Wunderlich2005,Sinova2015} on top of the magnetic insulators like YIG. These electrically created spin accumulations are usually detected by an indirect method called spin-Hall magnetoresistance effect in which the resistance of the NM changes with the magnetization of the underlying YIG \cite{Vlietstra_2013_PRB}. In these electrical measurements used to probe SHE, the ever present background contributions like Oersted fields and dipolar fields cannot be disentangled. Any technique that would aim to estimate these background contributions needs to be spin and magnetic field sensitive along with spacial resolution. 

Muon spin spectroscopy is widely used as a magnetic spin microprobe to investigate superconductivity\cite{Sonier2000,Kiefl2010}, magnetism~\cite{Dalmas2016,Guguchia2014} and many other fields~\cite{Zaher2014}. In addition, low-energy muon spin rotation spectroscopy (LE-$\mu$SR) provides an opportunity to tune the energy of the muons (1~-~30~keV) to perform depth resolved internal field measurements in range of 1~-~200~nm~\cite{Zaher2014,Morenzoni2003,Prokscha2008}. Due to the combination of sensitivity \cite{Guguchia2014} and the spatial resolution \cite {Zaher2014}, LE-$\mu$SR has been applied successfully to obtain the depth-resolved profile of the local magnetization in various thin films and heterostructures \cite{Suter_2004,Drew_2005}.
\begin{figure*}[ptb]
\includegraphics*[width=0.99\textwidth, clip]{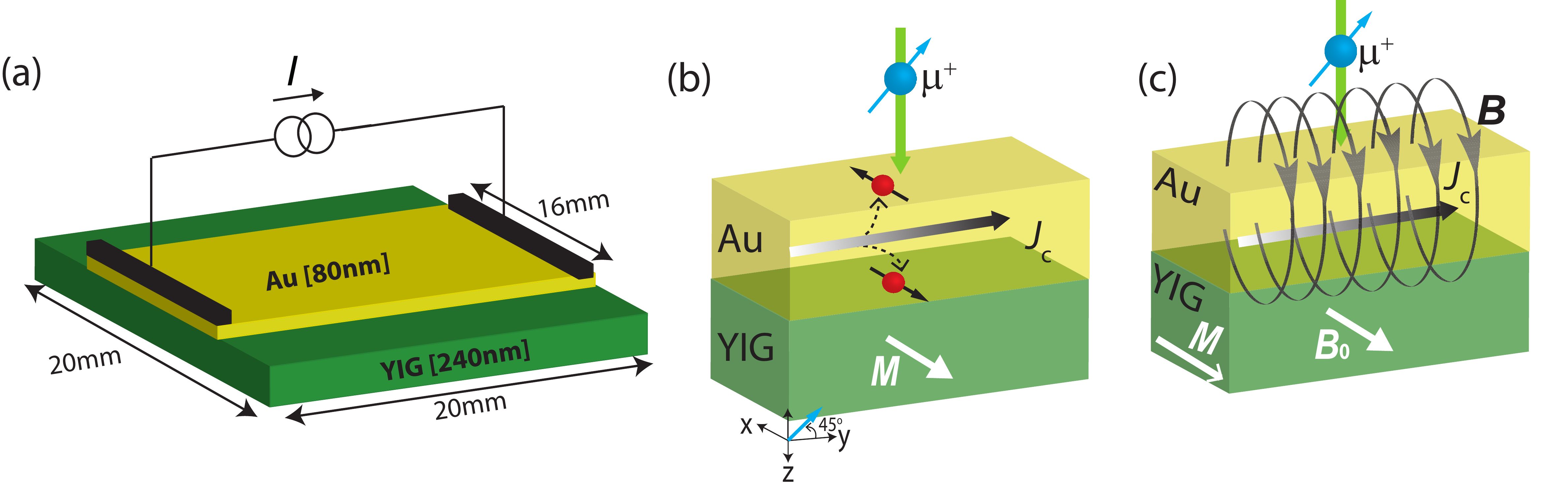}
\caption{\label{fig:1}
(a) Device configuration for probing current-induced magnetic fields at Au$|$YIG interface with muons. (b) Schematic illustration of spatial directions of electrically created spin accumulation created by spin-Hall effect and (c) Oersted magnetic fields $\bm B$  with respect to muon beam $\mu^+$. Here, $J_c$, $\bm M$ and $\bm B_0$ represent the applied dc-current, magnetization of the YIG film and the applied  magnetic field.}
\end{figure*}

All these successful application of LE-$\mu$SR motivates the study of its limits and capabilities in order to gauge the possibility of using such a technique for other sources of current-induced fields, e.g. due to the spin-accumulation by SHE, Oersted fields or magnetization induced via proximity at buried interfaces. To explore this, we considered here a Au$|$YIG test structure. In this structure, due to small spin-Hall angle of Au we expect negligible contribution from SHE, which allows us to quantify other current-induced contributions, such as ever-present Oersted fields.  We report here the quantitative study of depth distribution of magnetic fields in the Au$|$YIG system with LE-$\mu$SR \cite{Prokscha2006460, Bakule2004203}. 

Fig.~\ref{fig:1}(a) shows the device configuration used to quantify the current-induced magnetic field distribution at different depths in the Au$|$YIG heterostructure. The YIG has a thickness of 240~nm grown by liquid phase epitaxy on 0.5~mm thick (111) Gadolinium Gallium Garnet (GGG) single crystalline substrate. In any NM$|$YIG system, there would be two main contributions to a current-induced magnetization: one would be the spin accumulation due to SHE (see Fig.~\ref{fig:1}(b)) and other due to Oersted fields (see Fig.~\ref{fig:1}(c)). Note that for the Au metal (used here) we expect a spin diffusion length of 35~nm~\cite{Mosendz2010} which would make it compatible with depth-resolved studies of spin accumulation using LE-$\mu$SR. Nevertheless, for the specific case of SHE, the small spin-Hall angle makes the expected signals two orders of magnitude smaller than the Oersted fields, therefore in the current study we focus on quantifying the later. 

All measurements were performed at the LE-$\mu$SR spectrometers at the Paul Scherrer Institute, Villigen, Switzerland. In these measurements, 100$\%$ spin polarized positive muons are implanted into the Au$|$YIG sample. The implanted muons have a short life time of 2.2~$\mu$s after which they decay by emitting a positron, preferentially in the direction of the muon spin at the time of decay. The measurements reported here are done using the transverse field geometry, where the magnetic field is applied perpendicular to the initial spin direction of the implanted muons. The the decay positrons are detected using appropriately positioned detectors, to the left and right of the sample, relative to the incoming muons. The asymmetry, $A(t)$, in the number of detected positron in the left and right detectors (normalized by their sum) is proportional to the time evolution of the muon spin polarization, which provides information regarding the local magnetic properties at the muon stopping site. 

\begin{figure*}[tbp]
\includegraphics[width=0.99\textwidth, clip]{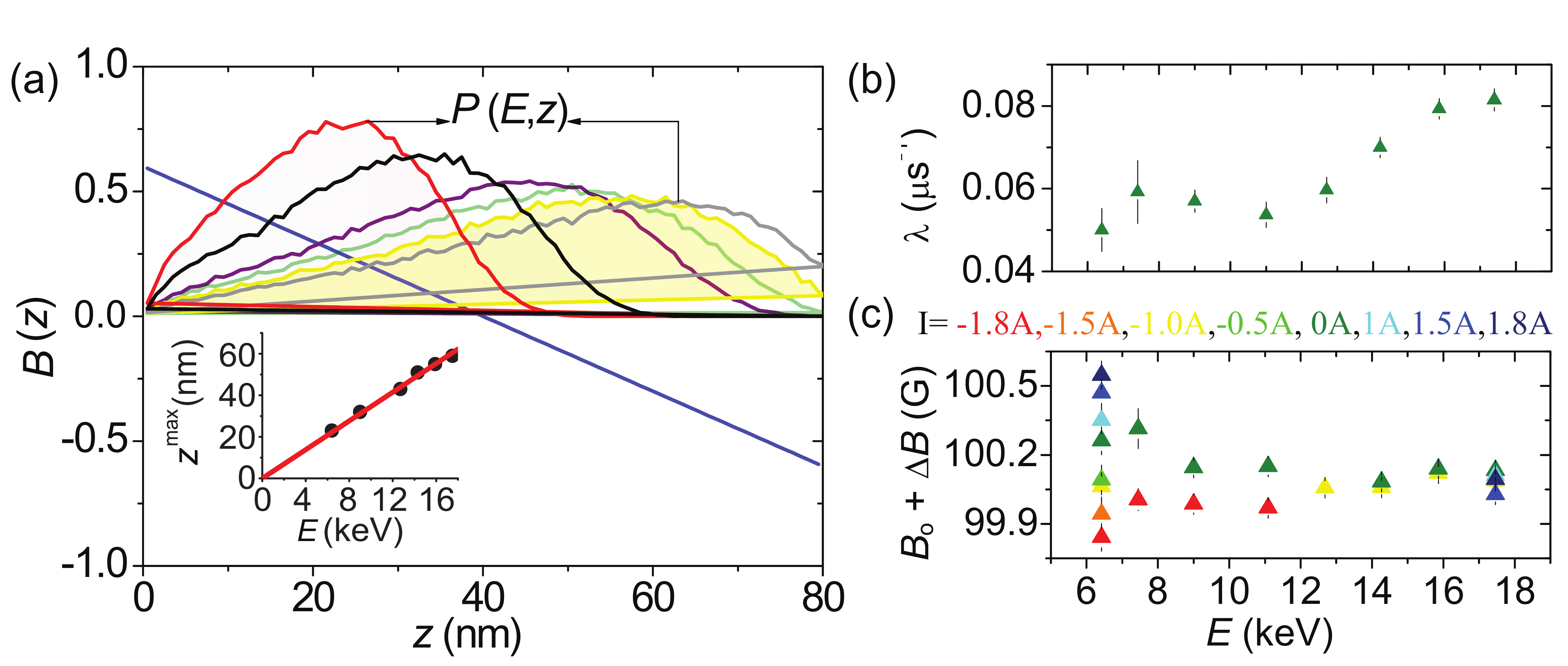}
\caption{\label{fig:2}
(a) The current-induced magnetic field as a function of depth $z$, where $z$ is the distance from the surface of Au towards the interface. $P(E,z)$ shows the probability distribution of the stopping depth of muons as a function of depth $z$ at different implantation energies $E$ varying from 6~keV to 18~keV. The inset of (a) shows the depth ($z^{\rm max}$) of the peak maxima for each probability distribution $P(E,z)$ shown in (a) versus $E$. It provides a scale ($z^{\rm max}=3.455 \times E$) to translate from $E$ to depth $z$. (b) and (c) shows the observed damping $\lambda$ and field $B_0+\Delta B$ as a function of the implantation energy $E$ of muons at different values of applied current (I =  -1.8~A to 1.8~A) in the Au$|$YIG bilayer system, respectively. Here, $B_0$ represents the applied field.
}
\end{figure*}

All measurements were done at pressure $\rm{\leq~10^{-9}}$~mbar in a cold finger cryostat. The magnetic field ($B_0= \rm 100~G$) was applied parallel to the Au$|$YIG interface along x-axis and muons were implanted with their spin polarization direction at an angle of $\rm{45^o}$ in the y-z plane, as shown in Fig.~\ref{fig:1}.
The depth profile of current-induced fields for Au$|$YIG are calculated by using $B = - \mu_0 J z$ with an assumption of $J$ as a uniform current density through the Au metal. By varying the energy of the muons, they can be implanted at different depths in the Au metal.  The implantation profiles of muons can be simulated using the Trim.SP Monte Carlo~\cite{Morenzoni2002}.  Fig.~\ref{fig:2}(a) shows the calculated stopping profiles  $P(E,z)$ of muons for this experiment as a function of distance $z$ from the Au surface. By tuning the implantation energy we can probe the magnetic properties closer (higher E) or further (lower E) from to the Au$|$YIG interface.

The measurements reported here were performed at different implantation energies and applied currents. The energies used and corresponding muons implantation stopping depth profile are shown in Fig.~\ref{fig:2}(a). These measurements allow us to probe the current-induced part of the magnetic fields as a function of distance from the Au$|$YIG interface. The obtained $\mu$SR spectra were analysed using the MUSRFIT software~\cite{Morenzoni2002}. We find that the collected spectra at all implantation energies and applied currents fit best to Eq.~\ref{eq3}.
\begin{equation}
A (t) = A_0 e^{-\lambda t} \cos(wt + \phi). \label{eq3}
\end{equation}
Here $\omega=\gamma B$, $\gamma$ being the muon gyromagnetic ratio, which reflects the fact the muons experience a Lorentzian field distribution with an average field B and width $\lambda$. The larmor frequency $\omega$ provides the information about the internal field at the muon site and the damping $\lambda$  gives information about the inhomogeneity of the internal field at the muon site.

The results of the fit parameters from Eq.~\ref{eq3} are shown in Fig.~\ref{fig:2}. For the damping $\lambda$ we do not observed any trend versus current therefore in Fig.~\ref{fig:2}(b) we show $\lambda$ only for zero current. Contrary to $\lambda$, there is a clear current dependence of the field shift $\Delta B$.  This dependence of $\Delta B$ is clearly larger at lower energies and gradually decreases until it fully disappears at higher energies ($E  \approx 12~\rm keV$) as shown in Fig.~\ref{fig:2}(c). When muons are implanted closer to the interface, we expect to observe an increase in $\Delta B$ due to the current-induced Oersted fields, as shown in Fig.~\ref{fig:2}(a). However, $\Delta B$ almost disappear closer to the interface. The internal field at the muon site is also measured at zero current density to rule out other magnetic field induced effects like proximity effects consistent with current understanding of the NM$|$YIG films~\cite{Nakayama2013}. 
\begin{figure*}[tbp]
\includegraphics[width=0.99\textwidth,natwidth=310,natheight=342]{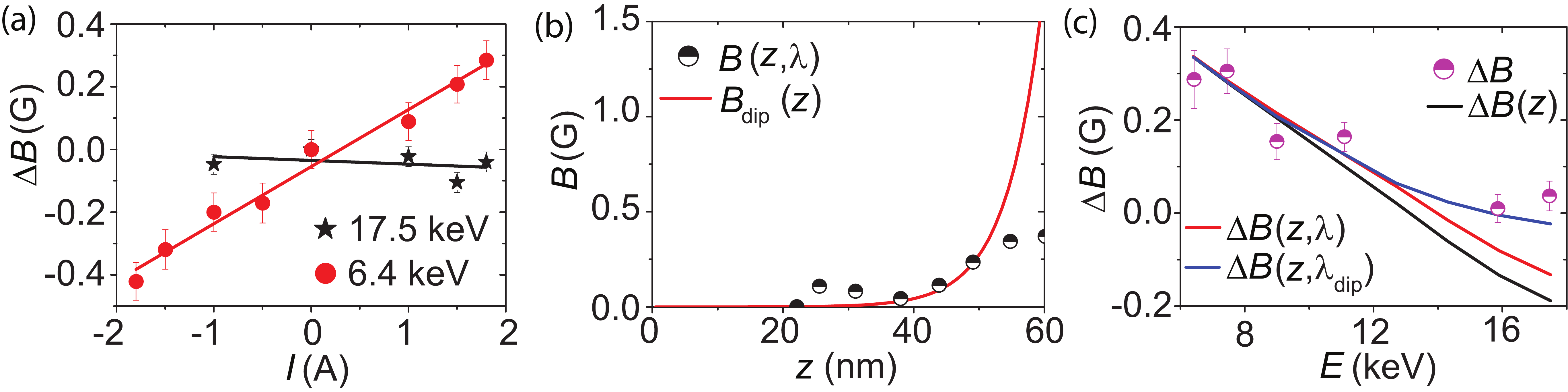}
\caption{\label{fig:5}
(a) Shift in the internal field $\Delta B$ at muon site as a function of the applied current $I$ through the Au film at energies $E=6.4~\rm keV,~17.5~\rm keV$ (b) Comparison between the calculated fields $B(z,\lambda)$ by considering the observed damping $\lambda$ shown in Fig.\ref{fig:2}(b) and fields $B_{\rm dip} (z)$ by considering the dipolar fields using Eq.~\ref{eq:3b} as a function distances $z$ from a ferromagnetic YIG surface. (c) Comparison of observed field shifts $\Delta B$ at $I=\pm 1.8 \ A$ with the calculated shifts $\Delta B(z)$, $\Delta B(z,\lambda)$ and $\Delta B(z,\lambda_{\rm dip})$ at different implantation energies of muons. Here, $\Delta B(z)$, $\Delta B(z,\lambda)$ and $\Delta B(z,\lambda_{\rm dip})$ represent the field shifts including, only the muon depth distribution profiles, the effect of observed damping $\lambda$ at $I=0$ and the effect of estimated damping $\lambda_{\rm dip}$ due to the dipolar fields, respectively. 
}
\end{figure*}
A clearer observation of the current dependence of $\Delta B$  for different energies is shown in Fig.~\ref{fig:5}(a). $\Delta B$ vary linearly with the applied current closer to the surface of the Au film at $E=6.4~\rm keV$ and almost vanishes closer to the interface at $E=17.5~\rm keV$, as shown in Fig.~\ref{fig:5}(a).

To understand the results shown in Fig.~\ref{fig:2}(b,c), we model the expected field shifts for different energies by taking into account the calculated muon depth profiles $P(E,z)$ shown in Fig.~\ref{fig:2}(a) as an initial approximation. 
\begin{equation}\label{eq:3}
\Delta B (z) = \frac{1}{\int_{z=0}^{z=d} P(E,z) dz} \int_{z=0}^{z=d}  P(E,z) B(z) dz.  
\end{equation}
Fig.~\ref{fig:5}b shows the average shift in the internal field  $\Delta B(z)$ at the muon site for different implantation energies, calculated by using Eq.~\ref{eq:3}. Fig.~\ref{fig:5}c shows that almost 0.2~G shift is expected to be observed at the interface, however the observed value is almost zero close to the interface. 

To understand such a large  discrepancy between the modelling (Eq.~\ref{eq:3}) and the observations, we expand this approximation by considering also the observed damping $\lambda (E)$ (shown in Fig.~\ref{fig:2}(b)) which increases by a factor of two closer to the Au$|$YIG interface. As the damping represents the inhomogeneity in the fields at the muon site, it can be understood that at the interface there is a broadening of the field distribution that smears the expected values due to the current-induced Oersted fields. This smearing therefore prevents to observe the expected field shifts versus current at the interface. These inhomogeneous fields $B(z,\lambda)$ can be understood to represent the smearing of fields coming from Lorentzian field distribution of width of $\lambda(z)/{2\pi \gamma}$ in mT. To find $\lambda(z)$, we started with the function $\lambda(E)$ and convoluted $\lambda(z)$ by taking $z$ to be the peak position $z^{\rm max}$ of the muon distribution at each implantation energy, as shown in the inset of Fig.~\ref{fig:2}(a). By using the convoluted $\lambda(z)$, we estimated the inhomogeneous fields $B(z,\lambda)$ resulting observed twice larger damping occurring at interface between magnetic and non-magnetic materials. The estimated  $B(z,\lambda)$ is around 0.3~G, which is in the same order as expected current-induced fields at the interface. If we consider that these fields are smearing out the real shifts $\Delta B$, we can take $\lambda$ as an inversely proportional effect and modify the Eq.~\ref{eq:3} as follows:   
\begin{equation}\label{eq:3a}
\Delta B(z,\lambda) = \frac{1}{\int_{z=0}^{z=d} \frac{P(E,z)}{\lambda (z)}  dz} \int_{z=0}^{z=d} \frac{P(E,z)}{\lambda (z)} B(z) dz.  
\end{equation}
Fig.~\ref{fig:5}(c) shows that these inhomogeneous fields closer to the interface results in preferential reduction of the shift around 30$\%$ close to the interface.

\begin{figure}[htbp]
\includegraphics[width=0.5\textwidth, clip]{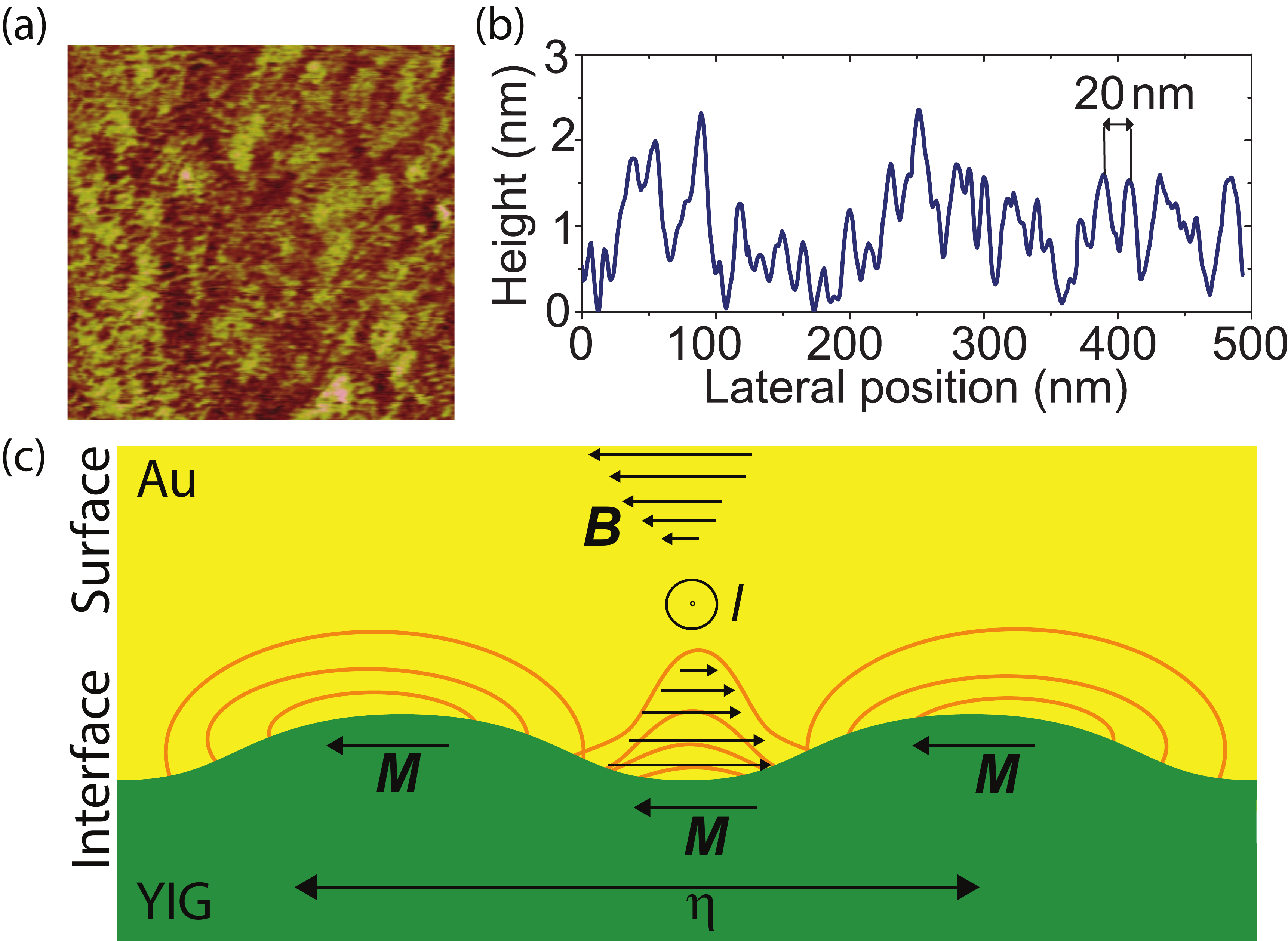}
\caption{\label{fig:6}
(a) Atomic force microscope image (500$\times$500~nm$^2$) and (b) a representative cross-sectional height profile of the YIG surface, prior to the Au metal deposition. (c) Illustration of inhomogeneous magnetostatic fields near the Au$|$YIG interface with finite roughness, sketched for a sinusoidal interface profile with a lateral period $\eta$. Here \textbf{M} and \textbf{B} represent the magnetization of YIG and the current-induced field, respectively. 
}
\end{figure}

There are several mechanisms that can influence the magnetic fields close to the interface including nuclear hyperfine fields~\cite{Merkulov2002,Dzhioev2002,Strand2003}, the dipolar fields from magnetic domains~\cite{Korenev2008} or the interface roughness~\cite{Aqeel2014,Dash2011}. The formers are not relevant here as nuclear hyperfine fields are too small in Au, typically 0.02~$\rm \mu s^{-1}$. We remark that the magnetic domains can be formed by anisotropy but for these films the anisotropy is not relevant and the thickness of YIG film is still small enough to neglect also the interfacial anisotropy, recently reported in thicker YIG films~\cite{Uchida2015}. However, the inhomogeneous magnetic fields arising from finite interface roughness can dramatically influence the dynamics of expected magnetic fields at the magnetic interface of multilayer systems~\cite{Dash2011}. The magnitude of these inhomogeneous dipolar fields scales with the roughness amplitude $h$ and decays with distance $z$ from the interface on a length scale of lateral roughness $\eta$~\cite{Demokritov1994,Dash2011}.  Fig.~\ref{fig:6}(c) shows the sketch of the dipolar fields near the Au$|$YIG interface with a finite roughness. The dipolar fields~\cite{Dash2011} can be estimated as follows:
\begin{equation}\label{eq:3b}
B_{\rm dip}(z) = \mu_0 M_s \frac{h}{2} \sum_{n=1}^{\infty} q_n \frac{\sin(\frac{1}{4}q_n \eta)}{\frac{1}{4}\eta q_n} \frac{\sin(\frac{1}{2}q_n h)}{\frac{1}{2}q_n h} \   {\rm exp}(- q_n z). 
\end{equation}
Here $q_n = \frac{2 \pi n}{\eta}$ and $M_s$ is the saturation magnetization of YIG. For this model of the sinusoidal interface profile, lateral period $\eta \rm =20~nm$ and roughness amplitude $h=1~\rm nm$ are estimated from the atomic force microscope image of the YIG surface shown in Fig.~\ref{fig:6}(a,b).  
Fig.~\ref{fig:5}(b) shows the dipolar fields $B_{\rm dip} (z)$ estimated by equation~\ref{eq:3b}. These $B_{\rm dip} (z)$ fields are larger than the inhomogeneuos fields $B(z,\lambda)$ estimated by considering observed damping at the interface which can be understood from the fact that $B(z,\lambda)$ are also convoluted from the muon profile $P(E,z)$, in reality we expect larger dipolar fields as estimated. 

To find the effect of these dipolar fields at the observed field shifts, we considered $\lambda_{\rm dip}$ associated with the dipolar fields as an inverse effect and used in equation~\ref{eq:3a}, as shown in Fig.~\ref{fig:5}(c). Fig.~\ref{fig:5}(c) shows a good agreement between the field shifts $B(z,\lambda_{\rm dip})$ estimated by including dipolar fields and the measured field shifts $\Delta B$, both vanishing closer to the Au$|$YIG interface. Therefore, we achieved a consistent picture by taking into account the damping $\lambda_{\rm dip}$ due to the dipolar fields resulting from the finite surface roughness. 

In conclusion we have established,  that LE-$\mu$SR can indeed work for resolving the background signals present due to interface roughness and Oersted fields which is universal feature in experiments done to probe SHE, with proper magnitude, distance dependence and  sign.  Our results serve as a guidance for future experiments aiming to probe SHE with muons.

We gratefully acknowledge J. Baas, H. Bonder, M. de Roosz and J. G. Holstein for technical support and funding via the Foundation for Fundamental Research on Matter (FOM), the Netherlands Organisation for Scientific Research (NWO), the Future and Emerging Technologies (FET) programme within the Seventh Framework Programme for Research of the European Commission under FET-Open Grant No. 618083 (CN-TQC), Marie Curie ITN Spinicur NanoLab NL, and the Zernike Institute for Advanced Materials National Research Combination. Part of this work is based on experiments performed at the Swiss muon source S$\mu$S, Paul Scherrer Institute, Villigen, Switzerland.


%
\end{document}